\documentstyle[prd, aps, floats, preprint, tighten, epsfig]{revtex}
\input{epsf}
\begin{document}
\onecolumn

\newcommand{\etal}{{\it et al.}}
\newcommand{\Dzero}{D\O}
\newcommand{\cdf}{CDF Collaboration}


\title{Application of Bayesian probability theory to the 
measurement of binomial data at past and future Tevatron experiments}

\author{Michael L. Kelly\footnote{Email address: 
              kelly@umaxp1.physics.lsa.umich.edu}}
\address{Department of Physics, University of 
              Michigan, Ann Arbor, Michigan, 48109-1120}

\date{August 10, 1999}
\maketitle

\begin{abstract}


The experimental problem of converting a measured binomial 
quantity, the fraction of events in a sample that pass a cut,
into a physical binomial quantity, the fraction of events 
originating from a signal source, 
is described as a system of linear equations. 
This linear system illustrates several familiar aspects of 
experimental data analysis. Bayesian probability theory is 
used to find a solution to 
this binomial measurement problem that allows
for the straightforward construction of confidence intervals.
This solution is also shown to provide an unbiased formalism for 
evaluating the behavior of data sets under different choices of 
cuts, including a cut designed to increase the significance of a
possible, albeit previously unseen, signal.

Several examples are used to illustrate the features of this method,
including the discovery of the top quark and searches for new particles
produced in association with $W^{\pm}$ bosons. It is also demonstrated
how to use this method to make projections for the potential discovery
of a Standard Model Higgs boson at a Tevatron Run 2 experiment, 
as well as the utility of
measuring the integrated luminosity through inclusive 
$p\overline{p} \rightarrow W^{\pm}$ production.
\end{abstract}



\section{Introduction}
\label{sec-intro}

A set of experimental data is almost always presented in terms of
the subset of interesting events, called the signal, and a complimentary subset
of events from non-interesting sources, called the background. 
The fraction of events from each subset is a binomial quantity; some fraction
of the sample is either characterized as signal or it is not. 
There is usually no exact means of
separating signal events from background events. Instead an experimental
cut is imposed on the original sample. Such a cut is 
motivated by independent studies that imply the cut will be more efficient 
for the events of interest than for the non-interesting events.
After this cut is applied, estimates are made regarding the amount of
signal and background in these new binomial subsets: those 
events that survived the cut, and those that failed the cut.

The attempt to rotate data from the experimental axis in 
pass-fail space onto the physics axis that defines 
signal-background space is referred to in this paper as 
the measurement problem. The measurement
problem is introduced and described in 
Section~\ref{sec-meas_prob}. The classical treatment of
the measurement problem as a system of linear equations 
provides some insight into the practical business of
analyzing data, but it is found to be inadequate for the
construction of confidence intervals. A Bayesian analysis
of related binomial quantities provides a straightforward
solution to this problem. 
Bayesian descriptions of binomial data are given
in Section~\ref{sec-tutorial}. A full solution to the
measurement problem is found in Section~\ref{sec-mysolution}.

The solution of Section~\ref{sec-mysolution} provides a
result in terms of the fraction of signal events in the entire
original sample; Section~\ref{sec-spass} reformulates this solution
so that the result can be presented as a fraction of signal
events in the subset of events that passed the cut.
The methods introduced in 
Sections~\ref{sec-mysolution} and~\ref{sec-spass} are
demonstrated in Example 1
with the data used for the discovery of the top quark.
Example 2 describes one way to use this method to estimate
the necessary size of control samples in order to 
understand the background to inclusive
$p\overline{p} \rightarrow W^{\pm}$ production.

Section~\ref{sec-discovery} presents a formalism for calculating 
the minimum number of events 
which must survive a cut designed to enhance the significance of
a possible signal over expectations.
Section~\ref{sec-xcheck} describes how to use the 
measurement problem to attribute a level of confidence
in the consistency of a possible new discovery with the original
understanding of the expected backgrounds.
Examples 3 and 4 illustrate the methods of 
Sections~\ref{sec-discovery} and~\ref{sec-xcheck}
using published 
$p\overline{p} \rightarrow W^{\pm} + b\overline{b}$ results.
Example 5 extrapolates this Tevatron Run 1 data to the estimated
amount of data available for a similar analysis in Run 2.

\newpage

\section{The Measurement Problem: A Linear System for Data Analysis}
\label{sec-meas_prob}

The measurement problem is equivalent to taking data that is
recorded on an experimental axis, {\it i.e.}~pass-fail space, and 
rotating the experimental results onto a physics axis, 
{\it i.e.}~signal-background space. When a cut is imposed on 
a data sample of 
$N_{Total}$ events, the sample is then divided into 
a subset of events which pass the cut $N_{pass}$, and a subset 
of events which fail the cut $N_{fail}$,
\begin{equation}  N_{Total} = N_{pass} + N_{fail} \; . \end{equation}
The original sample can also be described as a subset of signal 
$N_{sig}$ and background $N_{bkg}$ events,
\begin{equation}  N_{Total} = N_{sig} + N_{bkg} \; .\end{equation}
The different axes are related through a measurement matrix ${\bf M}$,
\begin{equation}
 \left( \begin{array}{c}  
         N_{pass} \\ 
         N_{fail} 
           \end{array} \right) =
{\bf M}
    \left( \begin{array}{c} 
           N_{sig} \\ 
           N_{bkg} 
            \end{array} \right) \; .
\end{equation}

The measurement problem is to invert the matrix ${\bf M}$ such that
\begin{equation} 
\left( \begin{array}{c} N_{sig} \\ N_{bkg} \end{array} \right) =
{\bf M}^{-1}
\left( \begin{array}{c} N_{pass} \\ N_{fail} \end{array} \right) ,
\end{equation}
where the elements of the measurement matrix are the efficiencies of
the cut on the signal and the background,
\begin{equation} 
{\bf M} \equiv
\left( \begin{array}{cc} \varepsilon & r \\
			1 - \varepsilon & 1 - r \end{array} \right) \; .
\end{equation}

The efficiency $\varepsilon$ of the cut on the signal  is 
defined as the number of signal events that will pass 
the cut, $S_{pass}$, divided by the total number of signal 
events in the original sample; the number of signal 
events which will fail the cut $S_{fail}$ is the total 
number of signal events times the inefficiency
($1 - \varepsilon$):

\begin{mathletters}
\begin{eqnarray} 
   S_{pass} & = & \varepsilon \: N_{sig} \; , \label{eq:signals} \\ 
   S_{fail} & = & (1 - \varepsilon) \: N_{sig} \; . 
\end{eqnarray}
\end{mathletters}

The efficiency $\varepsilon$
is always evaluated from some independent 
control sample of $\varepsilon_{Total}$ diagnostic events, 
where 
$\varepsilon_{pass}$ ($\varepsilon_{fail}$)
diagnostic events pass (fail) the cut;

\begin{mathletters}
\begin{eqnarray} 
    \varepsilon     & = & \varepsilon_{pass} / \varepsilon_{Total} \; , \\
    1 - \varepsilon & = & \varepsilon_{fail} / \varepsilon_{Total} \; .
\end{eqnarray}
\end{mathletters}

Similarly the efficiency of the cut on the 
background $r$, referred to as the `rfficiency', is defined 
as the number of background events that will pass the cut 
$B_{pass}$ divided by the total number of background events 
in the original sample, while the number of background
events which will fail the cut $B_{fail}$ is the total 
number of background events times the `inrfficiency' ($1 - r$):
\begin{mathletters}
\begin{eqnarray}
   B_{pass} & = & r \: N_{bkg}    \; , \\
   B_{fail} & = & (1 - r) \: N_{bkg} \; .
\end{eqnarray}
\end{mathletters}
Just as the efficiency is evaluated from an independent 
diagnostic sample, the rfficiency comes from another 
independent sample of $r_{Total}$ events
where $r_{pass}$ ($r_{fail}$)
diagnostic events pass (fail) the cut;
\begin{mathletters}
\begin{eqnarray} 
    r     & = & r_{pass} / r_{Total} \; , \\
    1 - r & = & r_{fail} / r_{Total} \; .
\end{eqnarray}
\end{mathletters}

It is common to refer to the 
rejection factor $R$
of a cut as the ratio of the different efficiencies
\begin{equation} 
   R \equiv \frac{\varepsilon}{r} \; , \label{eq:rej_factor}
\end{equation}
while the enhancement $E$ of a cut on a given sample can be defined as
\begin{equation} E \equiv \frac{\varepsilon - r}{r} 
                 = R - 1 \; .
\label{eq:enh_factor} 
\end{equation}

The inverse measurement matrix ${\bf M}^{-1}$ ,  
\begin{equation} {\bf M}^{-1} = \frac{1}{\varepsilon - r}
   \left( \begin{array}{cc}
          1-r & \; -r \\
          \varepsilon - 1 & \; \varepsilon
   \end{array} \right) \; ,
\end{equation}
exists only if the determinant of matrix ${\bf M}$ is not equal 
to zero, which is true whenever 
$\varepsilon \neq r$. 
This requirement 
is naturally satisfied whenever the rejection factor is not 
equal to one or the enhancement is non-zero. 
Usually a cut is chosen such that 
\begin{equation}
0 < r < \varepsilon < 1 \label{eq:erchoose} \; .
\end{equation}
There is nothing in this formalism which prevents the choice of a cut
such that $\varepsilon < r$; this is the situation where the
background is enhanced at the expense of the signal.

Once the inverse measurement matrix is known, it is possible 
to describe the number of signal (or background) events in terms
 of the number of events which pass (or fail) the cut:
\begin{mathletters}
\begin{eqnarray}
N_{sig} & = & \frac{(1-r) \: N_{pass} - r \: N_{fail}}{\varepsilon - r} 
              \nonumber \\
        & = & \frac{N_{pass} - r \: N_{Total}}{\varepsilon - r} 
              \label{eq:Nsig} \; , \\
N_{bkg} & = & \frac{(\varepsilon - 1 ) \: N_{pass} + \varepsilon \: N_{fail}}
                    {\varepsilon - r} \nonumber \\
	& = & \frac{\varepsilon \: N_{Total} - N_{pass}}{\varepsilon - r}
                       \nonumber \\
        & = & \frac{N_{fail} -  (1- \varepsilon) \: 
                 N_{Total}}{\varepsilon - r} \; .
\end{eqnarray}
\end{mathletters}

In fractional terms, defining
\begin{mathletters}
\begin{eqnarray}
f_{sig}  & \equiv & N_{sig} / N_{Total}  \; , \label{eq:fsig} \\
f_{bkg}  & \equiv & N_{bkg} / N_{Total}  \; , \\
f_{pass} & \equiv & N_{pass} / N_{Total} \; , \label{eq:fpass} \\
f_{fail} & \equiv & N_{fail} / N_{Total} \; , 
\end{eqnarray}
\end{mathletters}
then 
\begin{mathletters}
\begin{eqnarray}
f_{sig}  & = & \frac{f_{pass} - r}{\varepsilon - r} \; , \\
f_{bkg}  & = & \frac{f_{fail} - (1 - \varepsilon)}{\varepsilon - r} \; ,
\end{eqnarray}
\end{mathletters}
or
\begin{mathletters}
\begin{eqnarray}
f_{pass}  & = & (\varepsilon - r) \: f_{sig} + r \; , \label{eq:fpassasfsig}\\
f_{fail}  & = & (\varepsilon - r) \: f_{bkg} + (1 - \varepsilon) \; .
\end{eqnarray}
\end{mathletters}

The fraction of events which pass the cut $f_{pass}$ 
will always be found in the interval 
$0 \leq f_{pass} \leq 1$, and 
it is natural to restrict the fraction of signal events in 
the total sample $f_{sig}$ to the same interval. Practically this 
means that a physical solution to the measurement problem 
exists only if 
$r \leq f_{pass} \leq \varepsilon $. 
If the fraction of events 
that pass the cut is greater than the efficiency 
or less than the rfficiency, then the estimates of 
$\varepsilon $ and $r$ 
need to be reevaluated, as they are almost certainly incorrect. 

It is possible that there is more than one background present 
in the original sample, and that a cut has 
different rfficiencies for different backgrounds, {\it e.g.},
\begin{equation} \left( \begin{array}{c} 
        N_{pass} \\ N_{fail} \end{array} \right) =	
\left( \begin{array}{ccc}
	\varepsilon    & r_{1}    & r_{2} \\
	1- \varepsilon & 1- r_{1} & 1 - r_{2} 	
	\end{array} \right)
\left( \begin{array}{c} 
        N_{sig} \\ N_{bkg}^{(1)} \\ N_{bkg}^{(2)} \end{array} \right) \; .
\end{equation}
Such problems can always be reduced to the form
\begin{equation} 
\left( \begin{array}{c} N_{pass} \\ N_{fail} \end{array} \right) =
\left( \begin{array}{cc} \varepsilon & r' \\
			1 - \varepsilon & 1 - r' \end{array} \right)
\left( \begin{array}{c} N_{sig} \\ \sum N_{bkg}^{(i)} 
     \end{array} \right) \; ,
\end{equation}
where the total rfficiency $r'$ is the weighted sum of the individual rfficiencies,
\begin{equation} 
r' = \sum_{i=1} r_{i} \: f_{i} \; .
\end{equation}
The weights $f_{i}$ are the fractional amounts of the total 
background due to the individual backgrounds,

\begin{equation} 
f_{i} \equiv \frac{ N_{bkg}^{(i)} }
                     { \sum N_{bkg}^{(i)} } \; .
\end{equation}
This allows all problems to be reduced to the case of one 
signal source and one non-signal source, {\it i.e.}~one background.

The solution of the measurement problem can be approached 
from a purely algebraic viewpoint. If $\vec{e}$ is a vector representing 
the experimental basis, with pass and fail axes, and
$\vec{p}$ is the vector representing 
the physics basis, with signal and background axes, 
the measurement problem is written 
$\vec{e} = {\bf M} \: \vec{p}$. 
When 
there are uncertainties in either of the basis vectors, or in the 
measurement matrix, the measurement problem is written
\begin{equation} \left( \vec{e} + \delta\vec{e} \, \right) =
   \left( {\bf M} + \delta {\bf M} \right)
   \left( \vec{p} + \delta\vec{p} \, \right) \; .
\end{equation}
Finding the solution $\vec{p}$ with uncertainty 
$\delta \vec{p}$
is a classic problem in 
linear algebra. The uncertainty 
$\delta \vec{p}$ is known~\cite{condnum} to be limited:
\begin{equation} \frac{ \| \delta \vec{p} \, \| }{ \| \vec{p} \, \| }  \leq
   \frac{ \gamma({\bf M}) }{ 1- \gamma({\bf M}) 
                    \frac{\|\delta {\bf M} \| }{\| {\bf M} \| }} 
\left( \frac{\|\delta {\bf M} \| }{\| {\bf M} \| } +
       \frac{ \| \delta \vec{e} \, \| }{ \| \vec{e} \, \| } \right) 
  \label{eq:inequality} \; ,
\end{equation}
where  $\|\vec{p} \, \|$ 
denotes the norm of a vector (or matrix) and  $\gamma({\bf M})$ is a 
non-negative real scalar known as the condition number of 
the measurement matrix ${\bf M}$, 
\begin{equation} 
\gamma ({\bf M}) \equiv \| {\bf M} \| \: \| {\bf M}^{-1} \| \; .
\end{equation}

If $\gamma ({\bf M})$ is large, the measurement problem is said to be 
ill-conditioned. The condition number is equivalent for both the 
maximum absolute column sum (the $1$-norm) and the
maximum absolute row sum (the $\infty$-norm) of the measurement matrix,
subject to the constraints of Equation~\ref{eq:erchoose}:
\begin{equation} \gamma ({\bf M}) = \left\{ \begin{array}{lc}
		( 2 - (\varepsilon + r) ) / (\varepsilon - r) &
			\mbox{when $(\varepsilon + r) < 1$} \\
		(\varepsilon + r)  / (\varepsilon - r) &
			\mbox{when $(\varepsilon + r) \geq 1$} 
			\end{array} \right. \; .   
  \label{eq:def_gammaM}
\end{equation}

The only way to avoid a measurement matrix with a large 
condition number is to avoid $\varepsilon \approx r$. 
In other words, large rejection factors lead to 
better conditioned measurement problems; better conditioned measurement
problems lead to a smaller uncertainty $\delta\vec{p}$ in 
the quantities on the physics axis $\vec{p}$.

There are many different funtions of 
$\varepsilon$ and $r$ that can be offered as a statistic to weigh the
relative merit of one particular cut (with $\varepsilon_{1}$ and $r_{1}$)
versus another cut (with $\varepsilon_{2}$ and $r_{2}$).
Minimizing $\gamma ({\bf M})$ of Equation~\ref{eq:def_gammaM} is only one 
strategy that can be used to search for a `best' set of cuts. 
One other strategy may be to maximize the rejection factor $R$ of 
Equation~\ref{eq:rej_factor}. Another commonly encountered rule-of-thumb
is to maximize $\varepsilon \times R$; the extra factor of $\varepsilon$ 
is introduced to account for the fact that the amount of signal 
in the subset of $N_{pass}$ events is directly proportional to
$\varepsilon$, cf. Equation~\ref{eq:signals}.

\begin{figure}[ht!]
\centerline{\psfig{figure=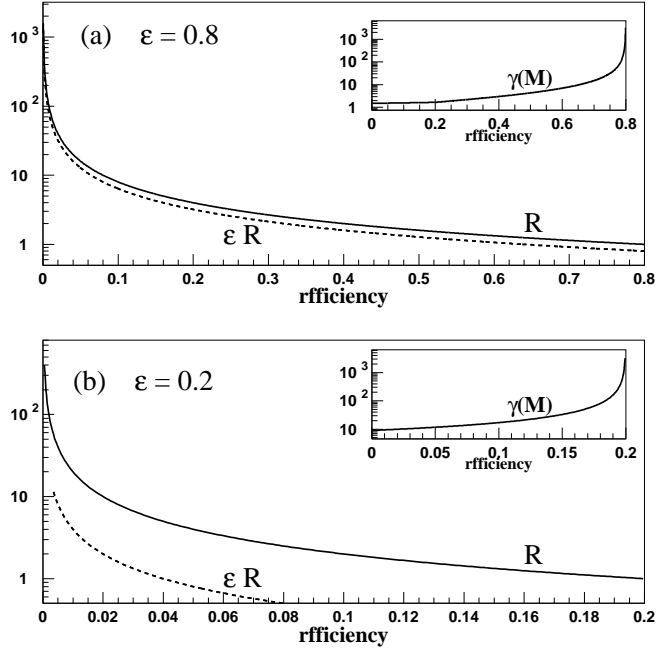,height=100mm}}
    \caption{The behavior of $\gamma ({\bf M})$, $R$, and 
$\varepsilon \times R$ as a function of $r$ for two different values
of efficiency $\varepsilon$. 
Possible strategies for the preference of one cut over another
is to maximize the rejection factor $R$ or to minimize the condition number of
the measurement matrix $\gamma({\bf M})$, see inserts. }
   \label{fig:erplot}
\end{figure}

Figure~\ref{fig:erplot} shows the behavior of these statistics for
the cases $\varepsilon = 0.8$ and $\varepsilon = 0.2$.
Each is undefined for cases of $r > \varepsilon$. Notice that strategies
that rely on minimizing $\gamma ({\bf M})$ see the 
relatively biggest improvement quickly as
$r$ takes values away from $\varepsilon$, but that a strategy of 
maximizing the
rejection factor sees the most improvement as 
$r$ approaches 0, independent of
the actual value of $\varepsilon$. 
As expected, both methods favor values of 
$\varepsilon$ closer to one than to zero.
None of these minimization or maximization strategies takes into account
any uncertainties in the state-of-knowledge of $\varepsilon$ or $r$, so
none of them can be considered an absolute statistic in deciding between
one cut or another.

While the linear algebra used to derive Equation~\ref{eq:inequality} 
provides some insight into the 
measurement problem, ultimately it
is unsatisfying in several respects. 
The most obvious limitation is that the upper limit on 
$\delta\vec{p}$ is 
not clearly defined in terms of confidence intervals. 
Another drawback is that there are several possible 
choices for the norm of the measurement matrix ${\bf M}$ 
and the uncertainty $\delta {\bf M}$. Yet another difficulty is the 
common confusion that arises from attempts to assign 
uncertainties to a binomial measurement, such as 
what fraction of events pass or fail a cut.

Bayesian probability theory provides a natural way to 
incorporate the knowledge, including the uncertainties,
of the efficiency, the rfficiency, and the measured experimental 
results $f_{pass}$ into a coherent statement about the state of 
knowledge of the physical signal fraction $f_{sig}$. 
Before describing 
the details of the solution to the measurement problem, 
the basics of Bayesian probability theory will be reviewed 
by considering its application towards binomial efficiencies.

\section{A Bayesian description of binomial data}
\label{sec-tutorial}

Bayesian probability theory (BPT) interprets a posterior 
probability density function (pdf) as the state-of-knowledge 
of an experimental result given some 
set of prior beliefs in the possible values of the result, 
the prior pdf, and the likelihood function describing the 
measured results of the experiment. The source of the 
posterior is Bayes' theorem:
\begin{equation} \mbox{posterior pdf} = \frac{ \mbox{prior pdf} \times 
                                \mbox{experimental likelihood} }
                              {\mbox{evidence} } \; .
\end{equation}
The maximum value of the posterior pdf is the most likely 
value, and the area beneath a particular interval along the 
posterior corresponds to the confidence that the true answer 
lies within the limits of that interval. 

\begin{figure}[ht!]
\centerline{\psfig{figure=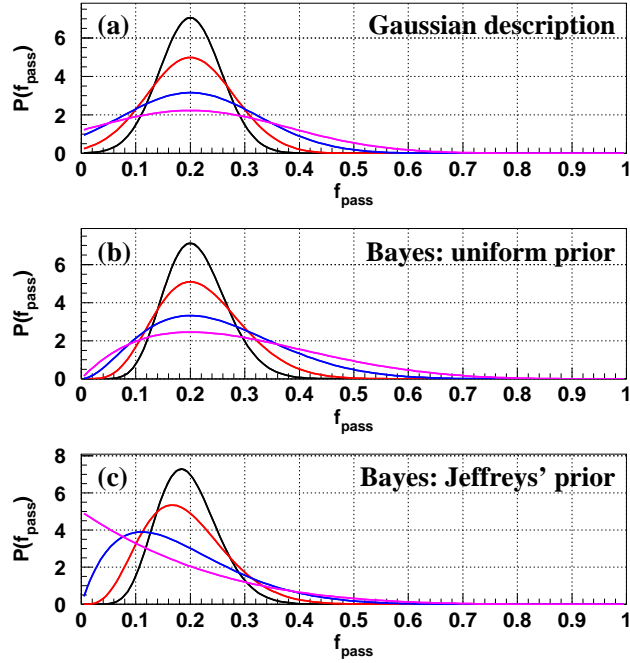,height=100mm}}
    \caption{Different posterior pdfs  from 
sample sizes of 5, 10, 20, and 50 events are shown.
Plot (a) shows normally distributed pdfs; only points within
the physical region are shown.
Plot (b) shows Bayesian posterior pdfs when a uniform 
distribution is used as the prior. The pdfs of (a) and (b) each
have a most 
likely value of $f_{pass} = 0.2$.
Plot (c) shows Bayesian posterior pdfs when 
Jeffreys' prior is used, where the fraction of events that
pass the cut is 0.2 for each pdf.
No part of any pdf in (b) or (c) extends beyond the physical region
$0 \leq f_{pass} \leq 1$.
}
   \label{fig:posteriors}
\end{figure}

A common problem encountered in the analysis of experimental results 
that is naturally described by BPT is the characterization 
of the uncertainties associated with the fraction of events that 
pass a particular cut $f_{pass}$. It has long been known
that binomially distributed quantities can be approximated by the
normal (Gaussian) distribution,

\begin{equation} P(f_{pass}; f_0, \sigma) = \frac{1}{\sigma \sqrt{2\pi}} 
   \exp \left( \frac{{\left( f_{0} - f_{pass} \right)}^{2}}{-2 \sigma^{2}} 
        \right) \label{eq:Gaussian} .
\end{equation}
The most likely value of the normal pdf is  
$f_{0} = N_{pass} / N_{Total}$ ,
and the variance is 
$\sigma = \sqrt{f_{0} (1-f_{0})/N_{Total}}$ . 
This approximation is only valid when 
$N_{Total} \gg 0$
and the mean  $f_{0}$ is not too 
close to the extreme values of one or zero. 
As $f_{0}$ approaches one or 
zero, the variance (as defined) approaches zero. It is not uncommon in 
experimental physics that one or both of these conditions is 
violated, leaving the approximation of Equation~\ref{eq:Gaussian} unusable.
In particular, experimental results often present
cases where zero events remain after a set of  
cuts is applied to a data sample. 
Figure~\ref{fig:posteriors}a shows normally distributed pdfs with the same 
$f_{0}$ but with different sample sizes $N_{Total}$. 
Note that the tails of the 
Gaussian distribution can extend beyond the physical 
region $0 \leq f_{pass} \leq 1$.

Several authors~\cite{gelman}~\cite{dagostini} have used the following 
posterior pdf to describe 
$P(f_{pass})$
in the interval $0 \leq f_{0} \leq 1$:
\begin{equation} P(f_{pass}; N_{pass}, N_{Total})
                  = \frac{ (N_{Total} + 1) \: !}
                                    { N_{pass}\:! \; 
(N_{Total} - N_{pass}) \: !} \:
{\left( f_{pass} \right)}^{N_{pass}} \:
{\left( 1 - f_{pass} \right)}^{N_{Total} - N_{pass}} \label{eq:DAP} \; .
\end{equation}
The most likely value for this posterior is simply 
the fraction of events that pass the cut, 
$f_{0} = N_{pass} / N_{Total}$. 
Figure~\ref{fig:posteriors}b shows 
Equation~\ref{eq:DAP} for different values of $N_{Total}$, each with the same 
most likely value of $f_{0} = 0.2 $.
The origin of this posterior is the use of a 
uniform prior over the physical region,
\begin{equation} P(f_{pass}) = \left\{ \begin{array}{rl}
                           1 & \mbox{if $0 \leq f_{pass} \leq 1$} \\
                           0 & \mbox{otherwise} \end{array} \right.  \; .
\end{equation}
The experimental likelihood is the binomial distribution,
\begin{equation} 
P(f_{pass}) = \frac{ N_{Total} \: !}{ N_{pass}\: ! \; 
(N_{Total} - N_{pass}) \: !} \:
{\left( f_{pass} \right)}^{N_{pass}} \:
{\left( 1 - f_{pass} \right)}^{N_{Total} - N_{pass}} \; .
\end{equation}
The evidence, or marginalization
term, normalizes the posterior to unit area, 
and is found by integration:
\begin{equation} \mbox{evidence} = 
\int_{-\infty}^{+\infty} \left( \mbox{prior} \times
                     \mbox{experimental likelihood} \right) df_{pass} \; .
\end{equation}
In the case of the uniform prior, the evidence term for this 
experimental likelihood distribution is equal to ${(N_{Total}+1)}^{-1}$.

It is possible to construct a different posterior 
pdf $P(f_{pass})$ with a different choice of prior, {\it e.g.},
\begin{equation} P(f_{pass}; N_{pass}, N_{Total})
             = \frac{N_{Total} \: !}
            { (N_{pass}-1) \: ! \; (N_{Total} - N_{pass}) \: !} \:
{\left( f_{pass} \right)}^{N_{pass} - 1} \:
{\left( 1 - f_{pass} \right)}^{N_{Total} - N_{pass}} \label{eq:Jeffreys} 
\end{equation}
arises if Jeffreys' (divergent) prior is used:
\begin{equation} P(f_{pass}) = \left\{ \begin{array}{cl}
                           1/f_{pass} & \mbox{if $0 < f_{pass} \leq 1$} \\
                           0 & \mbox{otherwise} \end{array} \right. \; .
\end{equation}
The evidence term in this case is equal to ${(N_{pass})}^{-1}$. The most 
likely value of Equation~\ref{eq:Jeffreys} is 
$f_{0} = (N_{pass} - 1)/(N_{Total} - 1)$, which approximates
the most likely values of Equations~\ref{eq:Gaussian} and \ref{eq:DAP}
as for large sample 
sizes. Figure~\ref{fig:posteriors}c shows the evolution of 
Equation~\ref{eq:Jeffreys} as the sample size is increased while the 
fraction of events that pass the cut is held constant.

The posterior pdf of Equation~\ref{eq:Jeffreys}
is included for completeness and will not be used in this
solution to the measurement problem. Notice that the use of
a divergent prior excludes cases of $N_{pass} = 0$.  
Jeffreys' prior would be used in those cases when an 
experimentalist claims complete ignorance of the efficiency
of the cut in the absence of any surviving events; {\it i.e.}, if
{\it zero} events pass the cut, the experimentalist 
who favors Jeffreys' prior
will not claim that {\it any} event will {\it ever} pass the cut. This is
obviously not satisfactory when attempting to set upper 
limits on a data sample with zero surviving events. In 
such a case, if the experimentalist is comfortable 
with setting the most likely value of the posterior 
$P(f_{pass})$ at $f_{0} = 0$ when $N_{pass} = 0$, a 
flat prior should be used.

\begin{figure}[ht!]
\centerline{\psfig{figure=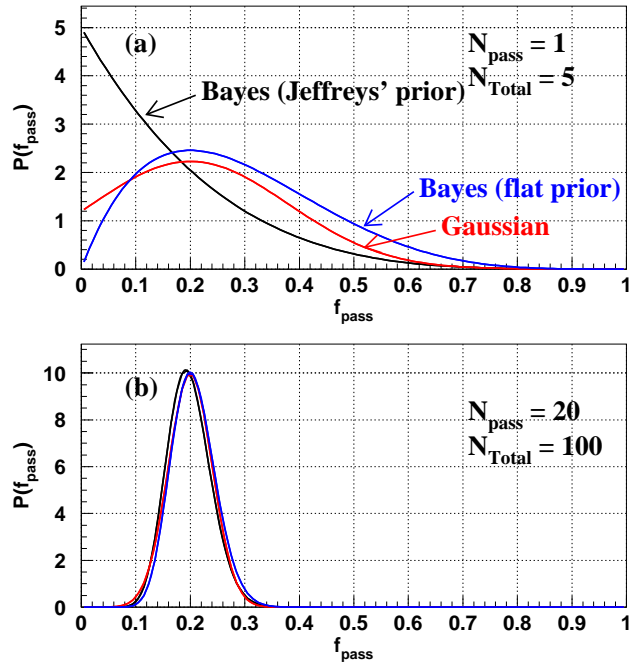,height=100mm}}
    \caption{The three different posterior pdfs for a sample 
of events where 20\% of the events pass the cut. Plot (a) shows the
posteriors for a sample size of five events. Plot (b) shows the
posteriors for a sample size of 100 events.
}
   \label{fig:three_nt}
\end{figure}

The three different posteriors, Equations~\ref{eq:Gaussian}, 
\ref{eq:DAP} and \ref{eq:Jeffreys}, demonstrate a 
feature of BPT; as the sample size increases the
posterior pdf becomes less sensitive to the particular 
choice of the prior pdf. Furthermore, as long as
the most likely value of the distribution
is not too close to its limiting values, as the sample 
size increases, the Gaussian distribution more 
closely approximates the posterior pdfs of 
Equations~\ref{eq:DAP} and \ref{eq:Jeffreys}. 
Figure~\ref{fig:three_nt}a shows the three different 
posteriors in the case of small sample size; Figure~\ref{fig:three_nt}b
shows the same distributions from a twenty times larger sample. 

\begin{figure}[ht!]
\centerline{\psfig{figure=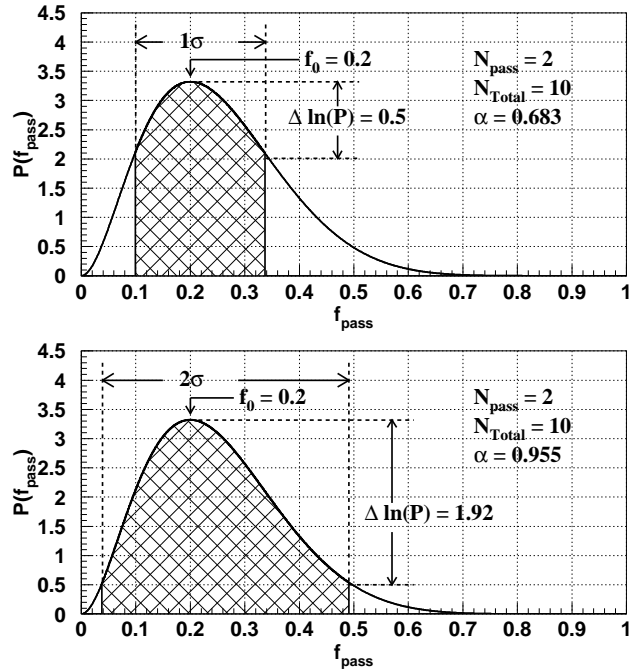,height=100mm}}
    \caption{The Bayesian posterior pdfs (uniform prior) for a 
sample of 10 events of which 2 events passed the cut. The 
hatched region of plot (a) represents the 0.683 confidence level interval
about the most likely value of $f_{pass} = 0.2$.
The hatched region of plot (b) represents the 0.955 
confidence level interval about the most likely value of $f_{pass} = 0.2$.
Also shown is the equivalence of these confidence intervals and
those defined by differences in the log-likelihood $\ln P(f_{pass})$.
}
   \label{fig:conf_level}
\end{figure}

\newpage
In this paper the notation 
$P(x)$ represents a pdf of no particular form. 
For complete generality, 
the pdfs for the efficiencies $\varepsilon$ and $r$
will be written as $P(\varepsilon)$ and $P(r)$;
it should be assumed that for the remainder of this paper each
is a shorthand representation for a binomial posterior described by
Equation~\ref{eq:DAP}.
For other cases, the notation 
$P(x; x_{pass}, x_{Total})$ will be used to describe a binomial
posterior of the form given by Equation~\ref{eq:DAP}, while 
$P(x; x_{0}, \sigma_{x})$
describes an explicitly Gaussian pdf of 
the form given by Equation~\ref{eq:Gaussian}.
Even though the posterior pdf represents the complete
knowledge of the particular distribution of possible 
values of a measured quantity, it is common to summarize 
the results of an experiment, {\it i.e.}~the posterior, with 
only a few numbers. Typically an experimental result will 
be quoted as the most likely value (the mode of the posterior)
with an upper and lower limit
such that the most likely value is contained
within the limits at some confidence level. For 
multi-modal posteriors it may be more appealing to quote 
the mean of the posterior rather than the modal value. 
Binomial problems do not of themselves give rise to multimodal posteriors.
For the purposes of this paper, the most likely value of a 
posterior will be quoted;
the most likely value of a posterior $P(x)$ will be represented $x_{0}$.
When error bars for a confidence 
level $\alpha$ are quoted, they describe the shortest interval about 
the most likely value that contains area $\alpha$
beneath the posterior pdf. Figure~\ref{fig:conf_level} shows the $\alpha = 0.683$
and $\alpha = 0.955$
confidence intervals for a Bayesian posterior constructed by 
Equation~\ref{eq:DAP}.

\section{A solution for the measurement problem}
\label{sec-mysolution}

In Section~\ref{sec-tutorial} it was shown that BPT can be used 
in binomial problems to construct posterior pdfs that exist 
completely within the allowed physical region. 
The application of cuts to finite data sets  
is a binomial problem: events in a data sample will either 
pass or fail a particular cut. While this classification is completely 
natural during the course of an experiment, the binomial 
quantity of interest is not the fraction of events which pass
 a cut, but the fraction of signal events in the original 
sample. A useful theorem of BPT provides a means of 
constructing a physical posterior pdf $P(f_{sig}$) from 
the experimental posterior pdf $P(f_{pass})$.

Generally, if a variable $y$ is a function of variable $x$, 
$y = f(x)$, an existing posterior distribution $P(y)$ can be 
used to construct a desired posterior function $P(x)$ simply 
by replacing $y$ in $P(y)$ with the functional form of $x$ and 
multiplying this posterior by the Jacobian~\cite{jacobian}:

\begin{equation} 
P(x) = 
  P\bigl(y(x)\bigr) \times \left| \frac{dy}{dx} \right| . \label{eq:changevars}
\end{equation}
In the measurement problem, this change of variables takes the form:

\begin{equation} 
P(f_{sig}) = 
    P\bigl(f_{pass}(f_{sig})\bigr) \times 
                \left| \frac{df_{pass}}{df_{sig}} \right| ,
\end{equation}
so that
\begin{equation} 
P(f_{sig}; \varepsilon , r, N_{pass}, N_{Total}) 
       = P \bigl( (\varepsilon -r) f_{sig} + r; N_{pass}, N_{Total} \bigr)
                \times | \varepsilon -r | \; .
\end{equation}

When Equations~\ref{eq:DAP} and \ref{eq:fpassasfsig} are used, 
the posterior pdf that describes the amount of signal in the 
original sample is:
\begin{eqnarray} 
P(f_{sig}; \varepsilon , r, N_{pass}, N_{Total}) & = &
    \frac{ | \varepsilon - r | \: (N_{Total} + 1) \: !}
	  { N_{pass}! \; (N_{Total} - N_{pass}) \: !} \times \nonumber \\
 &  &  {\Bigl( (\varepsilon -r) f_{sig} + r \Bigr)}^{N_{pass}} \:
{\Bigl( 1 - \bigl( (\varepsilon -r) f_{sig} + r \bigr) \Bigr)}^{N_{Total} - N_{pass}}
 \; .
\end{eqnarray}

The efficiency $\varepsilon$ and the rfficiency $r$ should be considered 
nuisance parameters as posterior pdfs 
$P(\varepsilon)$ and $P(r)$ 
can be constructed,
according to Equation~\ref{eq:DAP} in Section~\ref{sec-tutorial},
from independent control samples. Once 
these posteriors are known, 
{\it e.g.}~$P(\varepsilon; \varepsilon_{pass}, \varepsilon_{Total})$,
the nuisance parameters can be integrated away:
\begin{equation} P(f_{sig}; N_{pass}, N_{Total}) 
       = \int_{0}^{1} d\varepsilon \int_{0}^{1} dr \;
         P(\varepsilon) \; P(r) \;
         P \left( f_{sig}; \varepsilon , r, N_{pass}, N_{Total} \right)   
                \label{eq:mprob_solv}  \; .
\end{equation}

Equation~\ref{eq:mprob_solv} is the solution to the measurement problem. 
The efficiency $\varepsilon$, the rfficiency $r$, 
the size of the sample $N_{Total}$, and the number of events which 
pass the cut $N_{pass}$ all contribute to the posterior pdf for 
the fraction of the signal events in the original sample. 
This posterior is completely Bayesian: it will provide a 
most likely value for the signal fraction; it also allows 
for the natural construction of confidence intervals. 
Alternately, Equation~\ref{eq:mprob_solv} could be written in terms
of $P(f_{bkg})$, because of the trivial relationship
$f_{sig} + f_{bkg} = 1$.

This method of constructing 
$P(f_{sig}; N_{pass}, N_{Total}) $
from Equation~\ref{eq:mprob_solv} allows 
the state-of-knowledge of
$\varepsilon$ and $r$
to enter the understanding of $f_{sig}$ in a natural 
way. The efficiency and rfficiency originate from 
independent control samples; their most likely 
values depend on 
the particular cut used. It may be the case that 
$\varepsilon$ and $r$ 
have modal values $\varepsilon_{0}$, $r_{0}$
which lead to a measurement matrix with 
a small condition number, {\it q.v.} Equation~\ref{eq:def_gammaM},
but that they are found from 
such small diagnostic samples that the posterior pdfs 
$P(\varepsilon)$, $P(r)$ are 
very broad. This will lead to a broader distribution for 
$P(f_{sig})$ than the case of very precisely known
$\varepsilon$ and $r$.
Often experimentalists face the dilemma of diverting 
bandwidth from the recording of
possible signal sources to the task of
increasing the size of control samples,
especially when suffering from limited
statistics in one or more control samples. 
Equation~\ref{eq:mprob_solv} introduces an easy way to 
evaluate control 
samples of different sizes. See Example 2 below.

It should be noted that even in cases where 
$\varepsilon$ and $r$  are 
known very precisely, small sample sizes may lead to a 
pdf $P(f_{pass})$ which has non-zero values outside the 
physical region $r \leq f_{pass} \leq \varepsilon$. \
In such cases, the 
integral of $P(f_{sig})$  over the physically allowed values 
$0 \leq f_{sig} \leq 1$
 will be less than unity. It is useful to 
define an overall confidence level for the experiment,
\begin{equation} \alpha_{experiment} \equiv \int_{0}^{1} 
    P(f_{sig}; N_{pass}, N_{Total}) \; df_{sig} \label{eq:alpha_exp} \; .
\end{equation}

The value $\alpha_{experiment}$
is the confidence that the observed values of
$N_{pass}$ and $N_{Total}$
are consistent with the knowledge 
$P(\varepsilon)$ and $P(r)$. 
Equation~\ref{eq:alpha_exp} also defines the maximum confidence 
level that can be quoted for 
the posterior pdf $P(f_{sig})$. Since $P(f_{sig})$ is restricted 
to 
$0 \leq f_{sig} \leq 1$,
$(1 - \alpha_{experiment})$
is the fraction of the 
posterior that could not be constructed because 
part of $P(f_{sig})$ lies 
beyond the physical boundaries. 
Recall that by the construction of Equation~\ref{eq:DAP}, 
\begin{equation} 
   \int_{0}^{1} 
    P(f_{pass}; N_{pass}, N_{Total}) \; df_{pass} = 1 \; .
\end{equation}
While the Bayesian posterior pdf $P(f_{pass})$ is normalized to unity, 
representing complete certainty that the fraction of
events which pass a cut is between zero and one, the
posterior pdf $P(f_{sig})$ is not normalized, except by the 
Jacobian as seen in Equation~\ref{eq:changevars}.
A case of
$\alpha_{experiment}$ less than one
simply implies that larger data and/or diagnostic samples are 
required to increase the confidence 
that $f_{sig}$ is within the expected physical region. In cases 
where either
$\varepsilon$ or $r$ are
imprecisely known, the overall confidence 
of the experiment may be small if the fraction of events 
which pass the cut is very close to either 
$\varepsilon_{0}$ or $r_{0}$.  
In the event of an overall experimental confidence level of 
much less than one, the experimentalist is encouraged to alter 
the cut such that $f_{pass}$ is not too close to either
$\varepsilon$ or $r$,
or to increase the sample size $N_{Total}$.

Figure~\ref{fig:conf_level} shows the equivalence of 
Bayesian confidence level intervals and those constructed by
differences in the log of the posterior~\cite{pdgbook}.
In cases of 
$\alpha_{experiment} = 1$, the posterior is zero for
all values outside the physical region: here the log-likelihood
method will always provide an interval completely
within the physical
region. In the case of $\alpha_{experiment} < 1$, the
log-likelihood method may not be able to set one or
both limits within the physical region. See the insert
of Figure~\ref{fig:top_plot}b for Example 1 below for
an example of an experiment that could not set
bounds with a confidence level of greater than 93\%.

It should be noted that if for some reason the posteriors
$P(\varepsilon)$ or $P(r)$ are constructed by some formalism
other than that of Section~\ref{sec-tutorial} which causes 
one or both of them to be multimodal, the posterior 
$P(f_{sig})$ may become multimodal. This would be
a very unlikely circumstance that arises only through
the drastic intervention on the part of the experimenter.

This solution to the measurement problem
presents the results as a fraction of signal events $f_{sig}$
in the original sample of $N_{Total}$ events. It may be
preferrable for certain calculations, such as cross section measurements, 
to find the total number of signal events $N_{sig}$. 
It is trivial to use Equation~\ref{eq:fsig} to perform the change of
variables described by Equation~\ref{eq:changevars}:
\begin{equation}
	P(N_{sig}) = P \left( \frac{N_{sig}}{N_{Total}} ; N_{pass}, N_{Total} \right)
				 \left| \frac{1}{N_{Total}} \right | .
    \label{eq:changingfracs}
\end{equation}
The posterior $P(N_{sig})$ is defined on the interval 
$0 \leq N_{sig} \leq N_{Total}$.

Equation~\ref{eq:mprob_solv} is not only
a useful method for interpreting the results a single experiment,
but it can also be used as an unbiased tool to 
evaluate the possibility of applying different sets of 
cuts to the same original data sample, see Example 1 below. This
method has the further ability to quickly judge the possible 
improvements in a result from increased sample sizes, both for 
the data and control samples. Possible improvements may take the form
of a larger value of $\alpha_{experiment}$, a shorter confidence
level interval about the most likely value of $f_{sig}$, or both.

\newpage

\section{The measurement of signal in a subsample}
\label{sec-spass}

If an experiment is designed to extract a signal-rich sample 
of events by applying a cut, it is unusual to quote 
the fraction of signal events in the original sample that may be dominated
by background events. Rather 
than quoting the signal fraction from the $N_{Total}$ event sample,
it may be more useful to quote
either the signal fraction or 
the number of signal events in the subsample of $S_{pass}$ events 
that pass the cut. Recall that the equation for $S_{pass}$ is,
from Equations~\ref{eq:signals} and \ref{eq:Nsig},
\begin{eqnarray}
S_{pass} & = & \left( \frac{\varepsilon}{\varepsilon - r} \right)
              \left( N_{pass} - r \cdot N_{Total} \right) \nonumber \\
		 & = & \left( \frac{\varepsilon \cdot
			 N_{pass} }{\varepsilon - r} \right)
              \left( 1 - \frac{r}{f_{pass}} \right)  \;  .
\end{eqnarray}

Rearranging the above,
$f_{pass}$ can expressed as a function of $S_{pass}$:
\begin{equation} 
	f_{pass} = \frac{\varepsilon \cdot r \cdot N_{pass}}
	                {\varepsilon \cdot 
			N_{pass} - (\varepsilon - r) S_{pass} } \; .
	\label{eq:FpassasSpass}
\end{equation}
The number of signal events in the sample of events which passed the cut
is restricted to the interval $0 \leq S_{pass} \leq N_{pass}$. 
As in Section~\ref{sec-mysolution}, a posterior describing some
fraction of events $g$ will be used so that $0 \leq g \leq 1$.
The definition of $g$ for this problem is
\begin{equation}
	g \equiv \frac{ S_{pass} }{ N_{pass} } \; .
\end{equation}
 Equation~\ref{eq:FpassasSpass} is used to
express $f_{pass}$  as a function of $g$:
\begin{equation} 
	f_{pass} = \frac{\varepsilon \cdot r}
	                {\varepsilon - (\varepsilon - r) g } \; .
	\label{eq:FpassasG}
\end{equation}
Just as it was possible to change variables in order to construct 
a posterior pdf $P(f_{sig})$  from the form of the posterior for 
$P(f_{pass})$, it is possible to construct of $P(g)$.
The Jacobian from Equation~\ref{eq:FpassasG} is
\begin{equation}
	\left| \frac{d f_{pass}}{d g} \right| =
	  \left| \frac{\varepsilon \cdot r \: (\varepsilon - r)}
	                { { \bigl( \varepsilon - (\varepsilon - r) g \bigr) }^{2} }
		\right| \; .
\end{equation}
The posterior pdf $P(g)$ is then
\begin{equation} 
   P(g; \varepsilon , r, N_{pass}, N_{Total}) 
       = P \biggl(
                  \Bigl( \frac{\varepsilon \cdot r}
	                {\varepsilon - (\varepsilon - r) g } 
	              \Bigr) ; N_{pass}, N_{Total} 
			\biggr)
            \times 
			\left| \frac{\varepsilon \cdot r \: (\varepsilon - r)}
	                { {\bigl( \varepsilon - (\varepsilon - r) g  \bigr)}^{2} }  
			\right|  \; .
\end{equation}
As before, the nuisance parameters $\varepsilon$ and $r$
should be integrated away: 
\begin{equation} P(g; N_{pass}, N_{Total}) 
       = \int_{0}^{1} d\varepsilon \int_{0}^{1} dr \;
         P(\varepsilon) \; P(r) \;
         P \left( g ; \varepsilon , r, N_{pass}, N_{Total} \right)   \; .
		 \label{eq:preSpass_solv}
\end{equation}
Equation~\ref{eq:preSpass_solv} can be converted into a posterior
for $S_{pass}$ if a change-of-variables similar to that of 
Equation~\ref{eq:changingfracs} is performed, so
\begin{equation} P(S_{pass}; N_{pass}, N_{Total}) 
       = \int_{0}^{1} d\varepsilon \int_{0}^{1} dr \;
         {(N_{pass})}^{-1} \; P(\varepsilon) \; P(r) \;
         P \left( g ; \varepsilon , r, N_{pass}, N_{Total} \right)   \; .
		 \label{eq:Spass_solv}
\end{equation}
The posterior of Equation~\ref{eq:Spass_solv} 
is defined for the interval $0 \leq S_{pass} \leq N_{pass}$.
Similar posteriors can be constructed to describe $S_{fail}$, $B_{pass}$, or
$B_{fail}$.

\newpage

\subsection*{Example 1: The discovery of the top quark}

In 1995 the CDF~\cite{cdf_top_disc}
and \Dzero~\cite{d0_top_disc} collaborations reported 
conclusive evidence for
the process $p \overline{p} \rightarrow t \overline{t} + X$ production 
at the Fermilab Tevatron. The key to this discovery was the ability of
each experiment to devise cuts for the efficient selection of 
$t \overline{t}$ signal events from a parent sample of events with
a high transverse momentum lepton (from the decay of a $W^{\pm}$ boson)
and three or more jets. This important discovery can be used to
illustrate the use of Equation~\ref{eq:mprob_solv}.

The CDF experiment claimed discovery with 67 $\mbox{pb}^{-1}$ of integrated 
luminosity using analyses based on two different 
methods of discriminating signal from background through 
the identification of $b$ jets. The first method (SVX tag) 
identified $b$ jets by the reconstruction of secondary 
vertices within a silicon vertex detector. The second method (SLT tag) 
involved the reconstruction of soft leptons (here, electrons or muons) 
from the semileptonic decay of $b$ quarks. 
As expected these two methods had different 
efficiencies for both the signal and background, and different numbers of
events passed each cut. 

Table~\ref{tab:top_disc} contains the relevant information as published by
the CDF experiment. Also included is the CDF result~\cite{cdf_top_109} 
using SVX tags from a 
Run 1 data set corresponding to 109 $\mbox{pb}^{-1}$ of integrated 
luminosity. Even though the SVX and SLT
 methods have very different efficiencies, the
different methods show good agreement in the fraction of signal events in the
original sample of 203 $W^{\pm} + 3$ (or more) jets events. 
Notice
the agreement between the measured 
number of signal events $S_{pass}$ from
Equation~\ref{eq:Spass_solv} and the reported CDF numbers.

\begin{table}[ht!]
\caption{The discovery of the top quark at the CDF experiment:
Shown are the results of the solution to the measurement problem as 
applied to the published values of $\varepsilon$ and $r$.
Compare the solution's $S_{pass}$ to the published CDF value. 
Also shown are the CDF results from a Run 1 data set corresponding to
a larger integrated luminosity of 109 $\mbox{pb}^{-1}$.
}
\label{tab:top_disc}
\begin{tabular}{ccccccccc}

   & $N_{Total}$ & $N_{pass}$ & $\varepsilon$ (\%) & 
     $r$ (\%)& $f_{sig}$ & $S_{pass}$ &
	 $\alpha_{exp}$ & CDF  \\ \hline
	 
	SVX (67 $\mbox{pb}^{-1}$) &
		203 & 27 & $42 \pm 5$ & $3.3 \pm 0.1$ &
		$0.25^{+0.08}_{-0.07}$ & $22.5^{+2.3}_{-1.9}$ & 1.0 &
		$20.3 \pm 2.1$ \\

	SLT (67 $\mbox{pb}^{-1}$) &
		203 & 23 & $20 \pm 2$ & $7.6 \pm 0.1$ &
		$0.24^{+0.22}_{-0.18}$ & $13.2^{+4.6}_{-6.0}$ & 0.93 &
		$7.6 \pm 2.0$ \\
\hline
	SVX (109 $\mbox{pb}^{-1}$) &
		322 & 34 & $42 \pm 5$ & $3.3 \pm 0.1$ &
		$0.18^{+0.06}_{-0.05}$ & $26.0^{+2.3}_{-2.7}$ & 1.0 &
		$25.5 \pm 1.7$

\end{tabular}
\end{table}

\begin{figure}[ht!]
\centerline{\psfig{figure=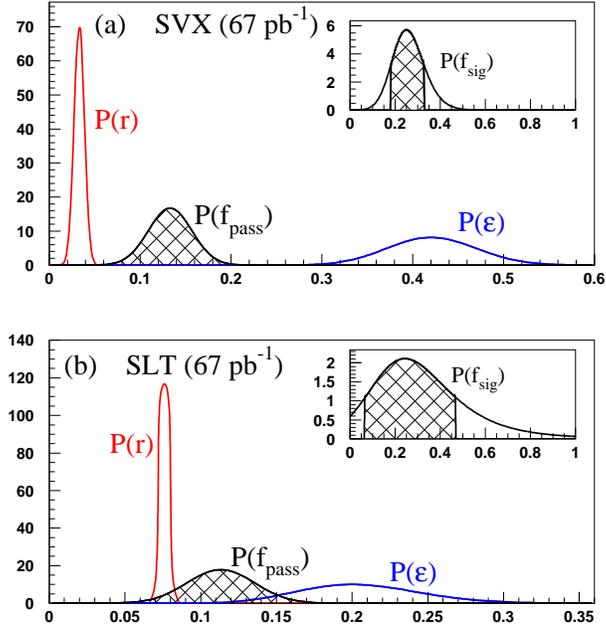,height=100mm}}
    \caption{The probability density functions of interest for CDF's
top quark discovery for the (a) SVX-tag sample, and (b) SLT-tag sample.
The inserts show the measured fraction of signal events in the entire
data sample; the hatched regions show the 0.683 CL interval about the
most likely value of $f_{sig}$.
Notice that the overlapping pdfs of (b) cause a less precise measurement of
$f_{sig}$.
 }
   \label{fig:top_plot}
\end{figure}

Figure~\ref{fig:top_plot} shows the state-of-knowledge of the 
efficiencies for signal and background, as well as the knowledge
of the fraction of events which survive the different $b$-tag requirements.
In Figure~\ref{fig:top_plot}a it is easy to see how the well-separated
pdfs $P(r)$, $P(\varepsilon)$, and $P(f_{pass})$ of the SVX measurement
give rise to a more precise measurement of $f_{sig}$. The overlapping
pdfs of the SLT measurement shown in Figure~\ref{fig:top_plot}b return a
similar most-likely value of $f_{sig}$, but the overall precision of the
SLT measurement is worse. The poor separation of the pdfs in the SLT
measurement leads to an overall experimental confidence level of 93\%,
compared to the 100\% confidence of the SVX measurement. This can be
seen graphically by comparing the inserts of Figure~\ref{fig:top_plot}:
The SVX result is completely within the physical region 
$0 \leq f_{sig} \leq 1$, while 7\% of $P(f_{sig})$ 
for the SLT result lies outside
the physical region.

\subsection*{Example 2: Using $W^{\pm}(\rightarrow e^{\pm} \nu)$ production as a luminosity
monitor at Tevatron Run 2}

It has been suggested~\cite{cdf_1a_wxsec}
that the two major Tevatron collider experiments count the
number of events from the process 
$p\overline{p} \rightarrow W^{\pm}(\rightarrow e^{\pm} \nu) + X$ during Tevatron Run 2
and use the theoretical predictions for the cross section times branching ratio
($\sigma_{signal}$) to measure the integrated luminosity (${\cal L}$).
It is hoped that such a `$W$ counting' method can measure the
integrated luminosity recorded at each experiment
more precisely than the approximately 4\% 
precision~\cite{cdf_1a_wxsec}~\cite{d0_wxsec}
used in the Run 1 physics results:
\begin{equation}
	{\cal L} = \frac{S_{pass}}
	                {\varepsilon \cdot A \cdot \sigma_{signal}} \; .
 \label{eq:lum_01} 
\end{equation}
The variables $S_{pass}$ and $\varepsilon$ have the same definitions as 
in Section~\ref{sec-meas_prob}; $A$ represents
the kinematic and geometric acceptance of the
detector used to collect the signal events. It is more natural to
reformulate Equation~\ref{eq:lum_01} in terms of the 
total number of signal events
recorded,
\begin{equation}
	{\cal L} = \frac{N_{sig}}{A \cdot \sigma_{signal}} \: ,
 \label{eq:lum_02} 
\end{equation}
so that the measured fraction of signal events is a function of the integrated
luminosity:
\begin{equation}
	f_{sig} = \frac{N_{sig}}{N_{Total}} 
        = \frac{ {\cal L} \cdot A \cdot \sigma_{signal}}{N_{Total}} \: .
 \label{eq:lum_03} 
\end{equation}
Following the derivation of Equation~\ref{eq:changingfracs}:
\begin{equation}
	P({\cal L}) = P \left( \frac{{\cal L} \cdot A \cdot \sigma_{signal}}{N_{Total}} ; 
						N_{pass}, N_{Total} \right)
				 \left| \frac{A \cdot \sigma_{signal}}{N_{Total}} \right | .
    \label{eq:lum_solution}
\end{equation}
The posterior $P({\cal L})$ is defined on the interval 
$0 \leq {\cal L} \leq {\cal L}_{max}$,  where
\begin{equation}
	{\cal L}_{max} = \frac{N_{Total}}{A \cdot \sigma_{signal}} \; .
 \label{eq:lum_max} 
\end{equation}

The variables of Equation~\ref{eq:lum_solution} are:
the number of total events ($N_{Total}$) prior to a selection of 
signal events ($N_{pass}$) by a cut with some efficiency ($\varepsilon$)
and rfficiency ($r$); the acceptance of the detector ($A$); and the 
theoretical cross section times branching fraction ($\sigma_{signal}$).
It will be assumed that the acceptance of a given detector can be known to
an arbitrarily small precision through the use of large Monte Carlo data sets
and a complete detector simulation, {\it i.e.}:
\begin{equation} P(A) = \left\{ \begin{array}{rl}
                           1 & \mbox{if $A = A_0$} \\
                           0 & \mbox{otherwise} \end{array} \right.  \; .
\end{equation}
The knowledge of the efficiency of the selection criteria, $P(\varepsilon)$,
will come from $Z^0 \rightarrow e^{+}e^{-}$ decays recorded during data taking; for 
Tevatron Run 2 the size of this sample will be twenty times the approximately 
five thousand such events recorded
during Run 1~\cite{d0_wxsec}.

Ignoring the uncertainty in the integrated luminosity, the background 
fraction in the sample of events which pass the cuts
($r \times f_{bkg}$) was the dominant source of uncertainty in the measured
$p\overline{p} \rightarrow W^{\pm}(\rightarrow e^{\pm} \nu) + X$ cross sections from
Run 1. It is impossible to predict the exact amount of background that a given 
experiment will have prior to data collection, so the most important experimental
question facing an experiment that wishes to use a process like inclusive
$W^{\pm}$ production is: 
{How much diagnostic data is needed to understand
the background to an arbitrary degree of accuracy?}

In order to perform such a study, it is necessary to assume that 
each Tevatron experiment 
will be delivered 2 $\mbox{fb}^{-1}$ of data, and that 
$\varepsilon$, $r$, and $A$ will be close to the currently reported values 
from Run 1. It will be assumed that there is no uncertainty in the 
theoretical cross section times branching ratio $\sigma_{signal}$; the value
of the reported \Dzero~measurement~\cite{d0_wxsec}
 will be used for $\sigma_{signal}$. 
The size of the total data sample
$N_{Total}$ will be the sum of the number of signal events ($N_{sig}$)
and an arbitrary number of background events ($N_{bkg}$) 
depending on the value of 
$r \times f_{bkg}$:
\begin{equation} N_{bkg} = \left( r \times f_{bkg} \right) 
		\frac{N_{sig}}{1 - (r \times f_{bkg})} \; .
\end{equation}
\begin{table}[ht!]
\caption{The assumed values for the Tevatron Run 2 
$p\overline{p} \rightarrow W^{\pm}(\rightarrow e^{\pm} \nu) + X$ cross 
section measurement at \Dzero~. 
The measured value of the Run 1b inclusive $W^{\pm}$ boson
production cross section times branching ratio 
from the \Dzero~experiment is used in place of
a theoretical value for $\sigma_{signal}$. 
The value of $\varepsilon_{Total}$ is an estimate of the
total number of 
$Z^0 \rightarrow e^{+}e^{-}$ events that will available
for the measurement of the efficiency.}
\label{tab:wxsec-common}
\begin{tabular}{cccccccc}

 Experiment  & $r_0$ & $\varepsilon_0$ & $\varepsilon_{Total}$ & 
     ${\cal L}$ & $A_0$ & $\sigma_{signal}$  \\ \hline
	 
	\Dzero &
		0.43 & 0.70 & 108 000 &
		2 $\mbox{fb}^{-1}$ & 0.465 & 2310 $\mbox{pb}$  
%

\end{tabular}
\end{table}

Table~\ref{tab:wxsec-common} lists the assumed values for the Run 2 
$W^{\pm}(\rightarrow e^{\pm} \nu)$
cross section measurement for the \Dzero~experiment. 
Table~\ref{tab:wxsec-result} shows the $1\sigma$ confidence limit in the measured
integrated luminosity that can be expected for a given experiment for different 
control sample sizes ($r_{Total}$). 
The QCD background fraction in the inclusive $W^{\pm}$ sample
is known to vary with instantaneous luminosity and trigger 
definitions~\cite{d0_wxsec}~\cite{d0_1a_wxsec};
Table~\ref{tab:wxsec-result}
shows the effect of different amounts of background on the precision of the 
luminosity measurement. 
Note that the figures for the \Dzero~experiment assume that
both the central (CC) and endcap (EC) calorimeters are used; if only the CC is
used the \Dzero~experiment can expect a more precise measurement 
of integrated luminosity since the background
fractions in the central region should be approximately one-fifth the value in 
the EC regions, even 
though 70\% of the acceptance for $W^{\pm}\rightarrow e^{\pm}\nu$ 
events is in \Dzero's central region.

It should be noted that the upgraded \Dzero~central tracker to be 
used in Run 2 will almost certainly have different values
of $\varepsilon_0$ and $r_0$ than used here. Nevertheless, 
Table~\ref{tab:wxsec-result} shows that even with moderately 
sized diagnostic samples (one-tenth
the size of the final $Z^0 \rightarrow e^{+}e^{-}$ sample) it should be
possible to measure the integrated luminosity to a precision of better than 1\%
with this method, assuming that the theoretical uncertainties can be kept at or 
below this level of precision.

\begin{table}[ht!]
\caption{Shown are the $1\sigma$ confidence level intervals about the
nominal Tevatron Run 2 integrated luminosity 
${\cal L}_0 = 2 \; \mbox{fb}^{-1}$, 
as a function
of the amount of background and the number of diagnostic events available to 
measure $P(r)$. 
}
\label{tab:wxsec-result}
\begin{tabular}{cccc}

 Experiment  & $r_0 \times f_{bkg}$ & $r_{Total}$ & 
     $1 \sigma$ interval about ${\cal L}_0$  \\ \hline
	 
	\Dzero~(nominal bkg) & 0.064 & 1 000   & $\pm \; 0.017 \;\mbox{fb}^{-1}$ \\
	\Dzero~(nominal bkg) & 0.064 & 10 000  & $\pm \; 0.012 \;\mbox{fb}^{-1}$ \\
	\Dzero~(nominal bkg) & 0.064 & 100 000 & $\pm \; 0.011 \;\mbox{fb}^{-1}$ \\
\hline
	\Dzero~(less bkg) & 0.030 & 1 000   & $\pm \; 0.013 \;\mbox{fb}^{-1}$ \\
	\Dzero~(less bkg) & 0.030 & 10 000  & $\pm \; 0.012 \;\mbox{fb}^{-1}$ \\
	\Dzero~(less bkg) & 0.030 & 100 000 & $\pm \; 0.011 \;\mbox{fb}^{-1}$ \\
\hline
	\Dzero~(more bkg) & 0.100 & 1 000   & $\pm \; 0.026 \;\mbox{fb}^{-1}$ \\
	\Dzero~(more bkg) & 0.100 & 10 000  & $\pm \; 0.012 \;\mbox{fb}^{-1}$ \\
	\Dzero~(more bkg) & 0.100 & 100 000 & $\pm \; 0.010 \;\mbox{fb}^{-1}$ 

\end{tabular}
\end{table}

\newpage
\section{The Confidence Level for the Possible Discovery of a Signal}
\label{sec-discovery}

It may be the case that an experiment provides a sample of $N_{Total}$ events, and
the expected number of events $\mu_{B}$ is modeled by a some pdf, 
{\it e.g.}~$P(\mu_{B}; B, \sigma_{B})$, where the most likely value $B$ is less 
than the observed number of events. In such a case, the 
experimentalist may believe that the excess, 
$N_{Total} - B$, 
is due to a real signal rather than a statistical fluctuation. 
The confidence level 
$\alpha_{excess}$
that the excess is due to something 
beyond the expected background can be defined~\cite{helene} as
\begin{equation} \alpha_{excess} \equiv
       1 - \!\! \sum_{N = N_{Total}}^{\infty} \int_{0}^{\infty}
         d\mu_{B} \; \frac{e^{-\mu_{B}} \: \mu_{B}^{N}}{N!} \; P(\mu_{B}) \; ,
\end{equation}
or
\begin{equation} \alpha_{excess} \equiv
         \sum_{N = 0}^{N_{Total}-1} \int_{0}^{\infty}
         d\mu_{B} \; \frac{e^{-\mu_{B}} \: \mu_{B}^{N}}{N!} \; 
		 P(\mu_{B})  \label{eq:alpha_excess_prime}\; .
\end{equation}
In the case of an excess, it should be assumed that $P(\mu_{B}) = 0$
for all $\mu_{B}> N_{Total}$, so Equation~\ref{eq:alpha_excess_prime}
can be written
\begin{equation} \alpha_{excess} \equiv
         \sum_{N = 0}^{N_{Total}-1} \int_{0}^{N_{Total}}
         d\mu_{B} \; \frac{e^{-\mu_{B}} \: \mu_{B}^{N}}{N!} \; 
		 P(\mu_{B})  \label{eq:alpha_excess}\; .
\end{equation}

It is natural to try to enhance the significance of a possible 
signal by reducing the expected background $\mu_{B}$. 
This is done by applying a 
cut on the sample of $N_{Total}$ events. A cut is usually chosen 
such that the rfficiency of the cut on the background events 
(here, the anticipated events described by $P(\mu_B)$ ) is 
small, while the expected efficiency of the cut on the possible signal 
is large. Equation~\ref{eq:Nsig} can be turned around;
\begin{equation} 
   \varepsilon \cdot N_{sig} = N_{pass} - r \cdot N_{bkg} \; .
\end{equation}
In order to claim a positive signal, there are two conditions. The trivial 
condition is that the efficiency $\varepsilon$ is non-zero. 
The more important condition is that 
$N_{pass} > r \cdot N_{bkg}$. 
If there is an excess, the 
significance of the excess in the sample of $N_{pass}$ events incorporates 
the knowledge of $r$ and $\mu_{B}$ through the pdfs $P(r)$ and $P( \mu_{B} )$:
\begin{equation} \alpha_{discovery} \equiv \sum_{N = 0}^{N_{pass}-1} 
           \int_{0}^{1} dr 
           \int_{0}^{N_{Total}} d\mu_{B} \;
               \frac{e^{-(r \cdot \mu_{B})} \: {(r \cdot \mu_{B})}^{N}}{N!} 
			   \; P( \mu_{B} ) \; P(r)  \label{eq:alpha_disc} \; .
\end{equation}

$N_{pass}^{disc}$ can be defined as the minimum 
number of events $N_{pass}$ from Equation~\ref{eq:alpha_disc} 
that guarantees a confidence level 
$\alpha_{discovery}$
in the signal. Just as in 
Equation~\ref{eq:mprob_solv}, the knowledge of the 
rfficiency $P(r)$ plays a critical part in the solution to this problem. 
The attempt to enhance a possible signal is a common exercise that 
is often fraught with difficulties; 
Equation~\ref{eq:alpha_disc} is an unbiased 
tool to estimate the minimum number of events which
must survive any new cut used to extract a possible signal. 

\newpage
\subsection*{Example 3: Discovery of a Higgs boson at CDF?}

In 1997 the CDF experiment~\cite{cdf_higgs}
reported a slight excess of events in the
process $p\overline{p} \rightarrow W^{\pm} + 2$ jet events
in events where the $W^{\pm}$ boson decayed to either an electron or
muon, and one of the jets in the event was identified
as coming from a $b$ quark decay by either the SVX or SLT tagging method.
The CDF data is summarized in Table~\ref{tab:cdf_wbb}.
\begin{table}[ht!]
\caption{The CDF results reporting a possible excess of
observed $W^{\pm} + 2$ jets sample where one or both of the jets has been
$b$-tagged. The significance of the excess ($\alpha_{excess}$) has
been calculated using a Gaussian distribution with mean $B$ and 
width $\sigma_B$. }
\label{tab:cdf_wbb}
\begin{tabular}{cccc}

 Sample & Number Observed & Background Estimate ($B \pm \sigma_B$) &
	$\alpha_{excess}$  
   \\ \hline
         
	$W^{\pm}$ + 2 jets (no tag)      & 1527 & -             & - \\
	$W^{\pm}$ + 2 jets (one tag)     & 36   & $30 \pm 5$    & 0.49 \\
	$W^{\pm}$ + 2 jets (both tagged) & 6    & $3.0 \pm 0.6$ & 0.90

\end{tabular}
\end{table}

An interesting but yet unobserved Standard Model process that has the 
experimental signature of a $W^{\pm}$ boson and two $b$-jets is 
associated production of $W^{\pm}$ boson and a Higgs boson~\cite{whiggs}, 
{\it i.e.}~$p \overline{p} \rightarrow W^{\pm} + H^0$.
It is reasonable to try and enhance the 49\% confidence level excess in
the CDF single-tag sample by requiring both jets to be tagged; 
Table~\ref{tab:cdf_wbb} shows that this requirement increases the confidence 
that the excess is due to a signal beyond expectations to 90\%.
Based on the CDF results, it is possible to estimate how many events would 
pass the double-tag requirement in order to claim a higher
significance. 
From the data shown in Table~\ref{tab:cdf_wbb} it
can be assumed that 
the rfficiency of the double-tag requirement is  
to be 10\%. Table~\ref{tab:Npass_discovery} shows the minimum 
number of events from the single-tag sample
that would have to pass the double-tag requirement in order to claim
excesses at the $2\sigma$ and $3\sigma$ levels. Also shown is the case where
only 5 events pass the double-tag rquirement, which has an
$\alpha_{excess}$ confidence level 
of only $1\sigma$. 
\begin{table}[ht!]
\caption{Shown are the minimum number of events $N_{pass}^{disc}$ 
that must pass a
cut in order to claim discovery of a signal at 
a given confidence level $\alpha_{discovery}$, for Example 3.}

\label{tab:Npass_discovery}
\begin{tabular}{ccccccc}

   $N_{Total}$ & $B$ & $\sigma_{B}$ & 
     $r_{pass}$ & $r_{Total}$ & $\alpha_{discovery}$ & 
	  $N_{pass}^{disc}$ \\ \hline
	 
	36 & 30 & 5 & 10 & 100 & 0.683 & 5 \\
	36 & 30 & 5 & 10 & 100 & 0.955 & 8 \\
	36 & 30 & 5 & 10 & 100 & 0.997 & 11 

\end{tabular}
\end{table}

\newpage
\section{Crosschecks for the possible discovery of a signal}
\label{sec-xcheck}

If after the application of the cut a 
significant excess is observed, it is straightforward 
to use the measurement problem
as a cross check of the possible discovery. In order to 
perform such a cross check, some knowledge 
$P(\varepsilon)$ of the 
efficiency $\varepsilon$ of the cut on the (possible) 
signal is necessary. If nothing is known 
about $\varepsilon$ except that it was 
large enough to provide discovery, {\it i.e.}~$\varepsilon > r$,
the uniform pdf prior to be used for $P(\varepsilon)$ is
\begin{equation} P(\varepsilon) = \left\{ \begin{array}{cl} 
                      0       & \mbox{when $\varepsilon \leq r$} \\
                      {(1-r)}^{-1} & \mbox{when $\varepsilon > r$}
                     \end{array} \right. \label{eq:edisc_uniform} \; .
\end{equation}

Otherwise, if there is some a priori knowledge assumed 
about the possible signal, there may be more informed 
knowledge of $\varepsilon$, perhaps from a Monte Carlo simulation. 
With the knowledge of the efficiency and the rfficiency for the cut, 
it is reasonable to assess the overall confidence 
that fewer than $N_{fail}^{disc}$ events will fail the cut, where
$N_{fail}^{disc}$ is the number of events which are removed
from the sample by a cut designed to enhance the possible signal.
The expected number of 
events that will fail the cut $\mu_{fail}$ is
\begin{equation} \mu_{fail} = (1 - \varepsilon) \;
     N_{Total} + (\varepsilon - r) \; \mu_{B} \; .
	 \label{eq:mufail}
\end{equation}
Recall that for a possible discovery $\varepsilon > r$, so
$\mu_{fail}$ will never be negative.
It is important to recognize that from the substitution of 
$N_{pass}^{disc} = N_{Total} - N_{fail}^{disc}$ into
Equation~\ref{eq:alpha_disc},
a small value of 
$N_{fail}^{disc}$ which leads to large value of 
$\alpha_{discovery}$ may be inconsistent with the original 
description of the expected background.

The confidence level $\beta$ that fewer than 
$N_{fail}$ events will fail the new cut
given the expected background $\mu_{B}$ is
\begin{equation} \beta = \sum_{N = 0}^{N_{fail}} 
           \int_{0}^{1} dr 
           \int_{0}^{N_{Total}} d\mu_{B} 
           \int_{0}^{1} d\varepsilon \;
               \frac{e^{-\mu_{fail}} \: \mu_{fail}^{N}}{N!} \;
               P(\varepsilon) \; P( \mu_{B} ) \; P(r)  \; \label{eq:beta} .
\end{equation}
The variable $\beta$ is used here because in cases where the 
efficiency and rfficiency are known very precisely, 
$\beta$ 
expresses the confidence that the initial description 
of the background distribution can still accommodate the new excess.
In other words, $\beta$ is 
a measure of how likely it is for $N_{pass}$ (or more) events 
to remain after the application of the 
new cut, given the assumption that the original sample
of events contains both a new signal and the expected background.

Notice that once $N_{Total}$ is fixed, the number of events 
$N_{pass}^{disc}$ necessary to claim a discovery with a confidence level 
$\alpha_{discovery}$ is a function of $r$ and $\mu_{B}$ only. 
Once a discovery is claimed, 
the confidence level $\beta$ in the original background description
depends not only on $r$ and $\mu_{B}$, but also $\varepsilon$. 
This reflects the fact that the efficiency $\varepsilon$ of a cut
has no meaning for a sample devoid of signal events.
In the limiting case of Equation~\ref{eq:beta}
where both $\varepsilon$ and $r$ approach unity with complete certainty,
$N_{fail}$ approaches zero and $\beta$ approaches an upper limit of 1; 
if no further cut is placed on the sample, there is complete confidence
that the original background description accomodates the observed
excess. 
A value of $\beta \approx 1$ implies that $P(\mu_B)$ can completely
accommodate an excess in $N_{pass}$; $\beta \approx 0$ implies that
the description of the background cannot accommodate such an excess.

\newpage
\subsection*{Example 4: The degree-of-belief for a Run 1 Higgs discovery}

The CDF data used in 
Example 3
is a case where a
double $b$-tag sample of $W^{\pm}$ + 2 jet events 
shows an excess of observed events over expectations
at the 90\% level, see Table~\ref{tab:cdf_wbb}.
For simplicity, and to insure that the mathematical description
of the number of background events never has a value
greater than the observed
number of events, the distribution of expected events $P(\mu_B)$
will be modeled here as
\begin{equation} P(\mu_{B}) = \left\{ \begin{array}{cl}
                           ({2 \: \sigma_B})^{-1} & 
	\mbox{if $B - \sigma_B \leq \mu_{B} \leq B + \sigma_B$} \\
                           0 & \mbox{otherwise} \end{array} \right.  \; .
\end{equation}
This model for the background lowers the significance of the excess in
the double-tag sample from 90\% to 87\%,
but it will serve as an approximation to a Gaussian description of the expected
background.
Three different cases will be considered for the possible new signal that
is causing the excess in the double-tag sample. The first possibility is that
the double-tag requirment has an efficiency of $\varepsilon_0 = 0.33$, which is the
reported efficiency for a $W^{\pm} + H^0$ signal in the Run 1 CDF detector. The second
possibility is that the efficiency is much higher, $\varepsilon_0 = 0.90$.
The final possibility is that there is no {\it a priori} knowledge 
about the nature of this new signal, the knowledge of 
the efficiency of the cut on the new signal is 
only that this efficiency 
is always greater than the rfficiency of the cut 
on the background, 
but that it has no preferred value, 
{\it q.v.}~Equation~\ref{eq:edisc_uniform}.

\begin{table}[ht!]
\caption{Shown is the confidence $\beta$ in the original
description of the background $P(\mu_B)$ assuming that 
$N_{pass}^{disc}$ events pass the cut with efficiency 
$P(\varepsilon)$. The first four rows assume a signal
 efficiency of 33\%; 
the second four rows assume an efficiency of 90\%. In these
cases $\varepsilon_{pass}$ and $\varepsilon_{Total}$
were chosen such that the width
of $P(\varepsilon)$ is approximately 15\%.
In the final four rows a uniform pdf 
for $P(\varepsilon)$ is assumed. The entries 
with $N_{pass}^{disc} = 6$ correspond to the observed CDF results.}

\label{tab:belief1}
\begin{tabular}{ccccccc}

   $\alpha_{discovery}$ & $N_{pass}^{disc}$ & 
	$\varepsilon_{pass}$ & $\varepsilon_{Total}$ &
	$\alpha_{experiment}$ & 
      $N_{fail}^{disc}$ & $\beta$ \\ \hline
	 
	0.683 & 5  & 33 & 100 & 0.76 &  31 & 0.57 \\
	0.868 & 6  & 33 & 100 & 0.84 &  30 & 0.50 \\
	0.955 & 8  & 33 & 100 & 0.84 &  28 & 0.36 \\
	0.997 & 11 & 33 & 100 & 0.58 &  25 & 0.18 \\
\hline
	0.683 & 5  & 90 & 100 & 0.77 &  31 & 0.77 \\
	0.868 & 6  & 90 & 100 & 0.87 &  30 & 0.72 \\
	0.955 & 8  & 90 & 100 & 0.97 &  28 & 0.59 \\
	0.997 & 11 & 90 & 100 & 1.00 &  25 & 0.39 \\
\hline
	0.683 & 5  & \multicolumn{2}{c}{uniform $P(\varepsilon)$}
 & 0.70 &  31 & 0.64 \\
	0.868 & 6  & \multicolumn{2}{c}{uniform $P(\varepsilon)$} 
 & 0.77 &  30 & 0.58 \\
	0.955 & 8  & \multicolumn{2}{c}{uniform $P(\varepsilon)$}
 & 0.82 &  28 & 0.45 \\
	0.997 & 11 & \multicolumn{2}{c}{uniform $P(\varepsilon)$}
 & 0.76 &  25 & 0.26 

\end{tabular}
\end{table}

Table~\ref{tab:belief1} shows the results of the cross checks for the
cases under consideration. The first part of 
the table 
describes the case where the efficiency of the possible signal is 
the value of the double-tag efficiency for $W^{\pm} + H^0$ production 
suggested by the
CDF analysis. The observed 
excess seen in the remaining 6 events has a statistical significance of
90\%. The likelihood that the original description of the background
($30 \pm 5$ events)
can accomodate a possible signal with such an efficiency is 50\%. 
Because any discovery must now include the description of the
signal efficiency $P(\varepsilon)$, 
there is a smaller overall confidence level 
in the new result, here 84\%, reflecting the fact that part of
$P(f_{pass})$ overlaps the pdfs $P(r)$ and  $P(\varepsilon)$.

\begin{figure}[ht!]
\centerline{\psfig{figure=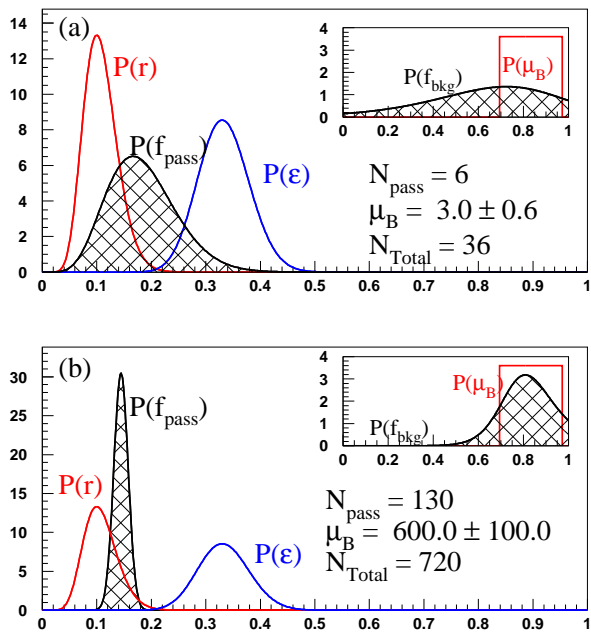,height=100mm}}
    \caption{Shown are the relevant pdfs for the cross check of a possible 
$H^0 \rightarrow b \overline{b}$ discovery at the CDF experiment
in associated $W^{\pm} + H^0$ production. Plot (a) shows
the fraction of the 36 events which pass the double 
$b$-tag requirement as a hatched
histogram superimposed upon the description of the efficiency for signal 
$P(\varepsilon)$ and background $P(r)$. The insert of (a) shows the pdf 
$P(f_{bkg})$, as a hatched histogram, that best describes the
observed number of events $N_{pass}$. This is to be compared with
the assumed description of the background $P(\mu_{B})$. 
Plot (b) shows the results for a hypothetical 
data set twenty times the size of the Run 1 result.
}
   \label{fig:cdf_wbb_plot}
\end{figure}

Figure~\ref{fig:cdf_wbb_plot}a is a graphical representation of
the cross check of the CDF result for the quoted efficiency of 
$\varepsilon_0 = $ 33\%.
The areas of $P(f_{pass})$ that overlap $P(r)$ and $P(\varepsilon)$ cause the
value of $\alpha_{experiment}$ to be less than unity. Although the insert 
of~\ref{fig:cdf_wbb_plot}a shows some 
agreement between the original background model $P(\mu_{B})$ and 
$P(f_{bkg})$ from Equation~\ref{eq:mprob_solv}; it is likely that the
original background description $P(\mu_{B})$ is too narrow, given the 
experimental results.

The rest of Table~\ref{tab:belief1} shows that it is more likely to see
a larger number of signal events if the efficiency of the cut 
on the new signal is larger, 
{\it i.e.}~90\% instead of 30\%. The last four lines 
of Table~\ref{tab:belief1} 
show that even if no knowledge about the efficiency of the 
possible new signal is claimed (except that the cut is more efficient 
for the signal that for the 
background), then the consistency $\beta$ of the initial 
background description to accommodate the observed 6 events is 58\%, 
slightly higher than if the double-tag signal efficiency is
described by a more precise, but overall smaller, value of 
$\varepsilon_0 = $ 33\%.

Tables~\ref{tab:belief1} 
illustrates some important trends
in attempts to increase the confidence level of a 
possible signal by imposing a further cut. 
The most important result is that knowledge of the cut
efficiency for the possible signal is useful but 
not necessary when trying to claim a discovery. 
For example, if the CDF collaboration chose to assume
a uniform pdf $P(\varepsilon)$ for the efficiency of the double-tag
cut on any new signal, the confidence level $\alpha_{excess}$ is 
still 77\%.
As 
in all measurement problems, a higher efficiency is 
better than a lower efficiency. This is 
important for increasing both
the overall confidence level of the experiment
$\alpha_{experiment}$ and the confidence $\beta$ 
that the new cut preserves the signal in a manner
consistent with
the original background estimate.

\subsection*{Example 5: Extrapolations for a Run 2 Higgs discovery}

The results of Examples 3 and 4 should not discourage anyone
from looking for a similar excess in a Tevatron Run 2 data set. It is trivial 
to scale the number of Run 1 events ($N_{Total}$) from Example 3
by the expected increase in integrated luminosity, 2 $\mbox{fb}^{-1}$ for Run 2,
and solve Equation~\ref{eq:alpha_disc} for the minimum number of events
$N_{pass}$ that must survive a double-tag requirement in order to have a
significant excess at some arbitrary confidence level.
Table~\ref{tab:run2_discovery} has the results for a twenty times larger
sample of events, assuming 
the same knowledge $P(\mu_B)$ and $P(r)$ used in Examples 3 and 4.
\begin{table}[ht!]
\caption{Shown are the minimum number of events $N_{pass}^{disc}$ 
that must pass the double-tag requirement
in order to claim discovery of a signal at 
a given confidence level $\alpha_{discovery}$, for Example 5.}

\label{tab:run2_discovery}
\begin{tabular}{ccccccc}

   $N_{Total}$ & $B$ & $\sigma_{B}$ & 
     $r_{pass}$ & $r_{Total}$ & $\alpha_{discovery}$ & 
	  $N_{pass}^{disc}$ \\ \hline
	 
	720 & 600 & 100 & 10 & 100 & 0.683 & 74 \\
	720 & 600 & 100 & 10 & 100 & 0.955 & 104 \\
	720 & 600 & 100 & 10 & 100 & 0.997 & 130 

\end{tabular}
\end{table}

Just as was done with the real data in Example 4, it is possible to test
the outcome of the Run 2 experiment with different assumptions for
the efficiency of the double-tag cut on any possible signal. 
The results are collected in Table~\ref{tab:run2_belief}. 
Figure~\ref{fig:cdf_wbb_plot}b plots the results of an experiment
with an assumed efficiency $\varepsilon_0 = $ 33\% for the signal where 130
events survive the double-tag cut, corresponding to a $3\sigma$ excess
in the original sample.
\begin{table}[ht!]
\caption{Shown is the confidence $\beta$ in the original
description of the background $P(\mu_B)$ assuming that 
$N_{pass}^{disc}$ events pass the cut with efficiency 
$P(\varepsilon)$. 
}

\label{tab:run2_belief}
\begin{tabular}{ccccccc}

   $\alpha_{discovery}$ & $N_{pass}^{disc}$ & 
	$\varepsilon_{pass}$ & $\varepsilon_{Total}$ &
	$\alpha_{experiment}$ & 
      $N_{fail}^{disc}$ & $\beta$ \\ \hline
	 
	0.683 & 74  & 33 & 100 & 0.46 &  31 & 0.81 \\
	0.955 & 104 & 33 & 100 & 0.86 &  28 & 0.51 \\
	0.997 & 130 & 33 & 100 & 0.98 &  25 & 0.23 \\
\hline
	0.683 & 74  & 90 & 100 & 0.46 &  31 & 0.97 \\
	0.955 & 104 & 90 & 100 & 0.86 &  28 & 0.88 \\
	0.997 & 130 & 90 & 100 & 0.98 &  25 & 0.76 \\
\hline
	0.683 & 74  & \multicolumn{2}{c}{uniform $P(\varepsilon)$}
 & 0.45 &  31 & 0.86 \\
	0.955 & 104 & \multicolumn{2}{c}{uniform $P(\varepsilon)$}
 & 0.81 &  28 & 0.66 \\
	0.997 & 130 & \multicolumn{2}{c}{uniform $P(\varepsilon)$}
 & 0.89 &  25 & 0.47 

\end{tabular}
\end{table}

The insert of Figure~\ref{fig:cdf_wbb_plot}b shows good agreement 
between $P(f_{bkg})$ and the broad $P(\mu_B)$ used in Example 4.
This implies that given the present knowledge of the efficiencies and 
the known processes, described by $P(\mu_B)$, that contribute events
to the $W^{\pm} + b\overline{b}$ sample,
it would not be surprising to observe a $3\sigma$ excess in
such a data sample from Run 2. The ultimate interpretation of
such an excess should come from an improved understanding of 
$P(\varepsilon)$, $P(r)$ and $P(\mu_B)$.

\newpage
\section{Conclusions}

The description of the measurement problem as a 
linear system of equations illuminates several 
important aspects of binomial experiments, including
intuitive notions about the relative value of
choosing cuts which preserve the signal of interest
while rejecting non-interesting backgrounds. 
The use of Bayesian techniques in this solution of the 
binomial measurement problem  offers a straightforward method of 
measuring signal and background fractions in both the
total data sample and the subset of events which survive
the application of a cut. It also 
provides an unbiased means of testing different  
cuts and is useful in evaluating potential improvements 
that come from increasing data samples. The method was 
also shown to be a powerful tool that can be used in the 
analysis of excess observed events over theoretical 
expectations.


%
%

%

\end{document}